\documentclass[journal]{IEEEtran}
\IEEEoverridecommandlockouts 
\usepackage{setspace, epsfig, psfrag, amssymb, amsmath, cite, color}
\usepackage{epstopdf,algorithm, algcompatible, lipsum, graphicx}
\usepackage{algpseudocode}
\usepackage{blindtext}
\usepackage{multicol}
\usepackage{soul}

\algnewcommand\INPUT{\item[\textbf{Initial Execution:}]}
\algnewcommand\OUTPUT{\item[\textbf{Repetitive Execution:}]}
\definecolor{m1}{cmyk}{0, 0.99, 0.4429, 0.3412} 
\definecolor{m2}{cmyk}{0, 0.61, 0.4429, 0.2412} 
\definecolor{m3}{cmyk}{.83, 0, .90, .35} 
\definecolor{m4}{cmyk}{.83, 0, .40, .11} 
\definecolor{m4}{cmyk}{.83, 0, .40, .11} 
\definecolor{m5}{cmyk}{0,0.5,1,0} 
\definecolor{myblue}{RGB}{0, 54, 114}
\def\bf{\normalfont\bfseries}
\def\bf{\normalfont\bfseries} 

\newfont{\smalll}{cmr8}
\def\IR{\mathbb{R}}

\def\IC{\hbox{C\hskip-
.5em\raise.5ex\hbox{$\scriptscriptstyle\mid$}}\ }
\def\Ic{\hbox{\smalll C\hskip-
.5em\raise.3ex\hbox{$\scriptscriptstyle\mid$}}\ }

\def\T={\buildrel {\scriptscriptstyle\triangle} \over =}

\def\sqr#1#2{{\vcenter{\vbox{\hrule height.#2pt\hbox{\vrule
width.#2pt height#1pt \kern#1pt\vrule width.#2pt}\hrule
height.#2pt}}}}

\def\block-diag{\mathop{\rm block{\scriptstyle -}diag}}

\def\pmbb#1{\setbox0=\hbox{#1}\raise 0.5ex\box0}

\newcommand{\bequ}{\begin{eqnarray}}
\newcommand{\eequ}{\end{eqnarray}}
\newcommand{\mT}{^\mathrm{T}}

\newcommand{\rom}{\mathrm}

\newcommand {\beq}      {\begin{equation}}
\newcommand {\eeq}      {\end{equation}}

\def\IR{{\mathbb R}}
\def\IC{{\mathbb C}}

\definecolor{tBlue}{RGB}{25,100,250}

\definecolor{tRed}{RGB}{250,5,40}

\usepackage{makecell}

\begin{document}
{
\title{{{\bf Orchestrated Couplings: A Time-Varying Edge Weight Framework for Efficient Event-Triggered Multiagent Networks}}}
} 

\author{Emre Yildirim$^\dagger$, Tansel Yucelen$^\dagger$, and Arman Sargolzaei$^\dagger$
\thanks{$^\dagger$E.~Yildirim, T.~Yucelen, and A.~Sargolzaei are with the Department of Mechanical and Aerospace Engineering at the University of South Florida, Tampa, FL 33620, USA (emails: {\tt\footnotesize emreyildirim@usf.edu, \tt\footnotesize yucelen@usf.edu, \tt\footnotesize a.sargolzaei@gmail.com}).}
}

\markboth{} {Shell \MakeLowercase{\textit{et al.}}: Bare Demo of IEEEtran.cls for Journals} \newcommand{\eqnref}[1]{(\ref{#1})}
\newcommand{\class}[1]{\texttt{#1}} 
\newcommand{\package}[1]{\texttt{#1}} 
\newcommand{\file}[1]{\texttt{#1}} 
\newcommand{\BibTeX}{\textsc{Bib}\TeX}
\maketitle



\begin{abstract}

In this paper, we focus on reducing node-to-node information exchange in distributed control of multiagent networks while improving the overall network performance. 
Specifically, we consider a multiagent network that is composed of leader and follower nodes over a time-varying, connected, and undirected graph.
In contrast to existing works on the event-triggered distributed control literature, we propose a time-varying edge weight event-triggered control framework.
In this framework, each node dynamically adjusts its edge weights by increasing them during the transient (active) phase and decreasing them during the steady-state (idle) phase of the multiagent network. 
This not only reduces the number of events in the network but also improves the performance (i.e., convergence speed and control effort)  of the overall multiagent network. 
System-theoretically, we first prove the closed-loop stability of the proposed event-triggered distributed control framework, where we then show that this framework does not exhibit a Zeno behavior. 
Finally, illustrative numerical examples are provided to demonstrate the efficacy of this framework.

\end{abstract}  


\section{Introduction}\label{introduction}
\subsection{Overview and Literature Review}
The popularity of multiagent networks is largely driven by their broad range of real-world applications including surveillance, agriculture, traffic and transportation, epidemics, and healthcare \cite{maldonado2024multi,afrin2021resource}.
In feedback control of multiagent networks, local information exchange is generally preferred over global (i.e., centralized) interactions. 
This preference arises from several limitations of global communication approach; that are, it lacks scalability due to communication costs, compromise security by exposing individual node objectives, prove impractical for large-scale network implementation \cite{cao2012overview,sargolzaei2024lyapunov}.
Consequently, the past two decades have witnessed substantial advances in distributed control architectures specifically designed to overcome these fundamental challenges while maintaining the performance and reliability of networks \cite{Lewis_book,mesbahi2010graph}.

Standard distributed control architectures generally rely on continuous or periodic information exchange between nodes \cite{Lewis_book,RenBeard:2008}. 
However, as network scale and operational complexity increase, 
these control architectures lead to data transmission bottlenecks (i.e., network congestion) and redundant energy usage \cite{nowzari2016distributed}.
To overcome these constraints, the event-triggered control theory introduces a solution by proposing aperiodic communication method \cite{tabuada2007event,dimarogonas2011distributed,heemels2012introduction}.
The event-triggered control theory relaxes the requirement of periodic information exchange.
However, it may result in significant deviations in their closed-loop multiagent network performance from their non-event-triggered counterparts due to fewer number of communication between nodes \cite{nowzari2016distributed,kurtoglu2024event}.

Several studies have already demonstrated that effective communication (i.e., the exchange of information) among nodes significantly influences the performance of the network, including convergence speed and the quality of collaboration \cite{olfati2007consensus}.
As mentioned in the previous paragraph, communication methods (i.e., periodic or aperiodic) also play a vital role in the network performance.
Furthermore, the number of neighboring nodes (i.e., connectivity) as well as the weights of the communication links (i.e., edge weights) affect the communication in the network \cite{krivzmanvcic2024adaptive}.
To this end, we seek an answer to the following scientific question: 
\noindent
\begin{center}
  \parbox{7.75cm}{%
    \centering
    \textit{Can we utilize time-varying edge weights as a control}\\ \textit{tool in such a way to improve the performance of the}\\  \textit{\hspace{-1 cm}network while reducing the number of events?}}
\end{center}

To the best of our knowledge, the majority of the existing works on event-triggered problems (see references in \cite{nowzari2019event,ge2018survey}) consider only constant couplings (i.e., edge weights) between nodes. 
On the other hand, there are limited number of studies  \cite{zhang2015distributed,jia2018consensus,jia2017event,zegers2021event} investigating the time-varying couplings. 
Specifically, the authors of \cite{zhang2015distributed} showed consensus of multiagent networks with event-triggered control in the presence of globally passive nonlinear couplings, and the authors of \cite{jia2018consensus} extended this result by using locally passive nonlinear couplings.
In addition, the authors of \cite{jia2017event} addressed consensus of multiagent networks with state-dependent nonlinear couplings.
The common denominator of these works is that they investigated time-varying couplings to model complex interaction between nodes.
However, the edge weights are not purposefully selected in these papers as a control strategy to enhance the performance of the network and reduce the number of events.
In contrast, \cite{zegers2021event} employed a reputation-based approach to assign time-varying edge weights within the range $[0,1]$ to mitigate the effects of Byzantine adversaries.
It is worth noting that the edge weights were chosen as either $0$ or a fixed positive value (not exceeding 1) in a switching manner.
Although, \cite{zegers2021event} suppressed adversaries by selecting the edge weights $0$ (i.e., cutting edges) with Byzantine adversaries and the network, they did not consider selecting different edge weights within non-adversarial nodes to improve the performance of the network and reduce the number of events, as presented in this paper.

Specifically, we propose a control strategy in which nodes dynamically adjust their edge weights according to the phase of the multiagent network over time, increasing them during transient (active) phases and decreasing them during steady-state (idle) phases. 
Note that the adjustment is not uniform across all edges; instead, edge weights are selectively increased, allowing the quality of information flow within the network. 
In the following subsection, we highlight the importance of edge weight selection and and demonstrate the benefits of allowing nodes to adapt their edge weights dynamically under transient (active) and steady-state (idle) phases.



\subsection{Proposed Control Strategy} \label{Intro_motivation}

\begin{figure}[t]
\vspace{0.2cm}
\hspace{0cm}
\centerline{\includegraphics[width=7.7 cm]{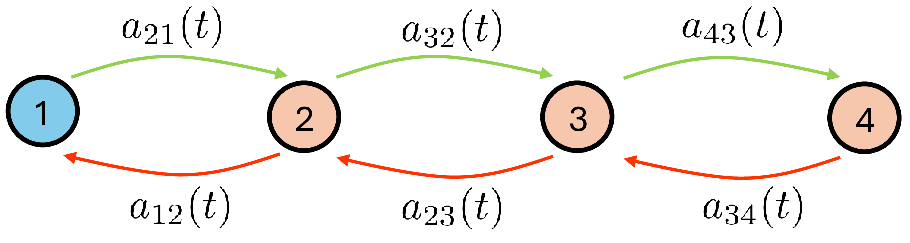}}
\vspace{-0.4cm}
\caption{Multiagent network composed of 4 nodes, where node 1 is the leader and the remaining nodes are followers. Color-coded edges given to illustrate the orchestrated couplings. Green edges (i.e., $a_{ij}(t))$ refer to larger edge weights comparing to their red counterparts (i.e., $a_{ji}(t)$) between nodes.}
\vspace{-0.3cm}
\label{graph0}
\end{figure}

Edge weights in multiagent networks quantify the relative importance of information flow between nodes \cite{leonard2024fast}.
While several studies \cite{dimarogonas2011distributed,kurtoglu2024event} in event-triggering control literature adopt uniform edge weight between nodes, this approach could be further improved by integrating time-varying edge weights to reduce the number of events and also increase the performance of the network.
To illustrate this point, consider the multiagent network in Figure \ref{graph0} as an example, where node 1 is the leader and the remaining nodes are followers.
Since only the leader holds the command information that all nodes aim to achieve, node 2 should assign a larger weight to the incoming edge (i.e., $a_{21}(t)$) from the leader than to the edge (i.e., $a_{23}(t)$) with node 3.
This prioritization\footnote{Each node decides its prioritization based on the number of updates re-
ceived from its neighboring nodes, which is further discussed in Remark 1.} ensures that node 2 places more emphasis on the information received from the leader than from node 3, thereby accelerating convergence and reducing the number of events.
Similarly, node 3 should set larger weights to the edge (i.e., $a_{32}(t)$) with node 2 than the one (i.e., $a_{34}(t)$) with node 4 since node 2 is closer to the leader node than node 4.
Hence, node 2 provides more up-to-date and accurate information to node 3.
This weighted framework naturally generalizes to the entire network, which improves the quality of information flow while mitigating the performance degradation caused by unnecessary information exchange.

Another important consideration regarding edge weights is their adaptation throughout different phases of the multiagent tasks. 
Multiagent networks typically operate in a transient (active) phase or steady-state (idle) phase.
For these periods, edge weights should adapt dynamically to these conditions.
In a transient phase, increased edge weights following the logic given in the previous paragraph enables rapid task completion with fewer events. 
Conversely, reduced influence between nodes in steady-state phases prevents unnecessary interactions and decreases event numbers.
This time-varying edge weight strategy provides better energy efficiency, fewer events, and superior multiagent network performance.
Although existing event-triggered control approaches consider time-varying edge weights, they neglect to coordinate these weights (i.e., orchestrated couplings) as described above to achieve better performance with fewer number of events.
Consequently, this paper goes beyond the scope of the existing event-triggered studies with time-varying couplings.



\subsection{Contribution}

The contribution of this paper is to address the scientific gap in the literature by proposing a time-varying edge weight event-triggered control framework. In this framework, each node dynamically adjusts its edge weights according to the phase of the multiagent network, increasing them during transient (active) phases and decreasing them during steady-state (idle) phases. We first establish the stability of the resulting event-triggered closed-loop network and show that it is free from Zeno behavior\footnote{Zeno behavior refers to the occurrence of an infinite number of triggering events within a finite time period \cite{nowzari2019event}.}. Two illustrative numerical examples are then provided to demonstrate the efficacy of the proposed method. Notably, the framework achieves superior performance compared to its constant edge weight counterpart without increasing the number of triggering events. Although similar performance could be attained by uniformly increasing constant edge weights, this approach incurs significant drawbacks, including a higher frequency of events, greater control effort, and more oscillatory state and control trajectories. 

\section{Mathematical Preliminaries}\label{notations}
In this paper, we use the standard notation.
Specifically, we respectively use $\IR$, $\IR^n$, and $\IR^{n \times m}$ for sets of real numbers, real vectors, and real matrices; $\IR_{>0}$ for the sets of positive real numbers; and ``$\triangleq$'' for the equality by definition. 
Moreover, $\text{diag}(a)$ is used to denote the diagonal matrix with the real vector $a \in \IR^n$ on its diagonal.
In addition, $[\hspace{0.07cm}\cdot\hspace{0.07cm}]^{-1}$ denotes the inverse and $[\hspace{0.07cm}\cdot\hspace{0.07cm}]\mT$ denotes the transpose.
Finally, $||\cdot||_2$ denotes the Euclidean norm for a vector and induced norm for a matrix.

With regard to the graph-theoretical notations, we consider a time-varying, connected, and undirected  graph $\mathcal{G}(t)\triangleq(\mathcal{V},\mathcal{E},\mathcal{A}(t))$, where $\mathcal{V}=\{\textbf{v}_1,\ldots,\textbf{v}_n\}$ is the nonempty finite set of $n$ nodes.
In addition, $(\textbf{v}_i,\textbf{v}_j)\in \mathcal{E} $ if there is an edge between nodes $\textbf{v}_i$ and $\textbf{v}_j$.
Moreover, $\mathcal{A}(t)=[a_{ij}(t)]\in{\mathbb{R}}^{n \times n}$  is the adjacency matrix of $\mathcal{G}(t)$, where $a_{ij}(t)>0$ if $(\textbf{v}_i,\textbf{v}_j)\in \mathcal{E}$, $a_{ij}(t)=0$ otherwise (self-loops are not allowed; that is, $a_{ii}(t)=0$, for $i\in \left\{1,\ldots,n\right\}$).
Furthermore, we define the neighbor set of each node $\textbf{v}_i$ as $\mathcal{N}_i = \{\textbf{v}_j\in\mathcal{V}|(\textbf{v}_i,\textbf{v}_j)\in \mathcal{E}\}$.
Next, we define the degree of node $\textbf{v}_i$ over a time-varying graph $\mathcal{G}(t)$ as $\text{d}_i(t) = \sum_{j\in \mathcal{N}_i} a_{ij}(t) $ for all $v_i \in \mathcal{V}$.
Let $\mathcal{D}(t)$ be  the degree matrix and $\mathcal{L}(t)$ be the Laplacian matrix, where $\mathcal{D}(t) = \text{diag}(\text{d}_1(t), \ldots, \text{d}_n(t))$ and $\mathcal{L}(t)=\mathcal{D}(t)-\mathcal{A}(t)$.
In addition, $\mathcal{L}(t)$ has zero row sums; that is, $\mathcal{L}(t)\textbf{1}_n=0$, where $\textbf{1}_n\in \IR^n$ is a vector of ones.



\section{Problem Formulation}\label{problem_formulation}
In this study, we consider a multiagent network composed of $n$ nodes over a time-varying, connected, and undirected graph $\mathcal{G}(t)$.
We categorize nodes into two groups; that are, leader nodes and follower nodes.
While leader nodes have access to the command, follower nodes do not.
Mathematically speaking, we express the dynamics of each node $i$ as
\begin{eqnarray}
\dot {x}_i(t)=u_i(t), \quad x_i(0)=x_{i0}, \label{node_dynamics}    
\end{eqnarray}
where $x_i(t)\in \IR$ denotes the state and $u_i(t)\in \IR$ denotes the control signal.
The control signal of node $i$ can then be given by
\vspace{0cm}
\begin{eqnarray}
{u}_i(t)\hspace{-0.05cm}&=&\hspace{-0.05cm}-\hspace{-0.05cm}\sum_{j\in \mathcal{N}_i} a_{ij}(t)\bigg((x_{\rom{s}i}\hspace{-0.01cm}(t)-x_{\rom{s}j}\hspace{-0.01cm}(t)) \nonumber\\ && + k_i (x_{\rom{s}i}\hspace{-0.01cm}(t)-c(t))\bigg).\label{control_signal}   
\end{eqnarray}
In (\ref{control_signal}), $c(t)\in \IR$ denotes the time-varying command and $x_{\rom{s}i}\hspace{-0.01cm}(t)\in \IR$ denotes the latest state broadcast (i.e., $x_i(t)$) of node $i$ to its neighboring nodes with event-triggering.
Moreover, $k_i=1$ if node $i$ is leader and $k_i=0$ otherwise.

The event-triggering rule adopted in this study is given as
\begin{eqnarray}
|x_i(t)-x_{\rom{s}i}\hspace{-0.01cm}(t)|\leq \delta, \label{outer_event_rule} 
\end{eqnarray}
where $\delta\in \IR_{>0}$ is the triggering threshold for broadcasting the updated states to the neighboring nodes.
In particular, once the trigger condition in 
(\ref{outer_event_rule}) is violated for node $i$ at time $t=t_{di}$, then $x_{\rom{s}i}\hspace{-0.01cm}(t)$ is set to $x_i(t_{di})$.
Then $i^\text{th}$ node broadcasts its latest state (i.e., $x_{\rom{s}i}\hspace{-0.01cm}(t)$) to its neighboring nodes.
In addition, $x_{\rom{s}i}(t)$ stays constant by neighbors of node $i$ using a zero-order-hold operator over the interval $t\in [t_{di},t_{(d+1)i})$, until node $i$ transmits an updated $x_{\rom{s}i}(t)$ to its neighbors following the next violation of its event-triggering rule at time $t=t_{(d+1)i}$.

In this paper, we consider time-varying edge weights (i.e., $a_{ij}(t)$) in the network.
Each node dynamically adjusts their edge weights.
To formalize this dynamic edge weight update mechanism (i.e., orchestrated couplings), we first present the following condition
\begin{eqnarray}
|x_{\rom{s}i}\hspace{-0.01cm}(t)-x_{\rom{s}j}\hspace{-0.01cm}(t)|\leq \delta, \quad \forall j\in \mathcal{N}_i. \label{data_recording}
\end{eqnarray}
Once (\ref{data_recording}) is violated, then node $i$ begins monitoring the frequency of state broadcasts from neighboring nodes as this likely signals the start of another mission (i.e., active phase). 
During this monitoring period, node $i$ records\footnote{This recording is later used in (\ref{gamma_update_rule}) to select control coefficient $\gamma_{ij}(t)$, which governs the prioritization of edge weights $a_{ij}(t)$. This provides efficient information flow throughout the network. Hence, the performance of the network improves as well as the number of event decreases.} the number of state broadcasts of its all neighboring nodes (i.e., $\phi_{ij}(t))$.
We note that if (\ref{data_recording}) is no longer violated, then node $i$ stops recording any broadcast counts of neighboring nodes and resets them to zero.
In addition to (\ref{data_recording}), we introduce the second condition for nodes to adjust their edge weights, which is given by
\vspace{0cm}
\begin{eqnarray}
|x_{\rom{s}i}\hspace{-0.01cm}(t)-x_{\rom{s}j}\hspace{-0.01cm}(t)|\leq \epsilon, \quad \forall j\in \mathcal{N}_i,\label{inner_triggering} 
\end{eqnarray}
where $\epsilon\in \IR_{>0}$ denotes the triggering threshold for edge weight update.
When (\ref{inner_triggering}) is violated, then edge weights of node $i$ are increased.
Otherwise, they are decreased.
The key away is that nodes increase their edge weights to execute the task relatively fast,  which yields fewer event-triggering actions.
On the other hand, they decrease their edge weights to minimize each other's influence when there is no task to accomplish (i.e., idle phase), which also reduces the number of events.

Once nodes decide whether to increase or decrease their edge weights based on (\ref{inner_triggering}), the edge weights are dynamically updated according to 
\begin{eqnarray}
\dot{a}_{ij}(t) &=&-\zeta_1(a_{ij}(t)- \theta_{ij}(t)), \quad a_{ij}(0)=a_{ij0}, \label{aij_dynamics}\\
\dot{\theta}_{ij}(t) & = & -\zeta_2 (\theta_{ij}(t)-\rho_{ij}(t)), \quad \theta_{ij}(0)=\theta_{ij0}, \label{thetaij_dynamics} 
\end{eqnarray}
where $\zeta_1\in \IR_{>0}$ and $\zeta_2\in \IR_{>0}$ are design parameters and $\theta(t)\in \IR$ is a low-pass filter of edge weights $\rho_{ij}(t)$ satisfying
\begin{eqnarray}
\rho_{ij}(t) =
\begin{cases}
 a_\text{min}, \quad \ \quad \quad \text{if} \ (\ref{inner_triggering}) \ \text{holds},  \\
\gamma_{ij}(t)a_\text{max}, \quad \text{otherwise}.
\label{edge_weight_dynamics}
\end{cases}
\end{eqnarray}
Here, $a_\text{min}\hspace{-0.05cm}\in\hspace{-0.05cm} \IR_{>0}$ and $a_\text{max}\hspace{-0.05cm}\in\hspace{-0.05cm} \IR_{>0}$ denote minimum and maximum edge weights in the network, respectively.
In addition, $\gamma_{ij}(t)\in \IR$ denotes a time-varying control coefficient.


Note that the condition (\ref{inner_triggering}) determines whether the edge weights should be increased or decreased.
However, it lacks differentiating the relative importance of edges, which plays a vital role in the network performance and the number of events as highlighted in the first paragraph of Section \ref{Intro_motivation}. 
This prioritization is handled by condition (\ref{data_recording}) and $\gamma_{ij}(t)$.
The following remark outlines the edge weight selection process.

\textbf{Remark 1.} Note that $\gamma_{ij}(t)$ in (\ref{edge_weight_dynamics}) governs the prioritization of edge weights in the active phase by using (\ref{data_recording}).
Specifically,
if (\ref{inner_triggering}) is violated, all edge weights of node $i$ increase, though not all reach $a_\text{max}$.
The key away is that node $i$ increases edge weights to enhance the overall network performance, while assigning relatively lower weights by using $\gamma_{ij}(t)$ to certain edges helps filter out irrelevant information from specific nodes.
To illustrate this point, consider the multiagent network in Figure \ref{graph0}.
If the states of all nodes in the network stay close enough to the command $c(t)$ (i.e., idle phase), then no nodes broadcast its updated state to their neighboring nodes since (\ref{outer_event_rule}) is not violated.
Once the leader node (i.e., node $1$) receives updated command $c(t)$ (i.e., active phase), then it starts to broadcast its state information to its neighboring node (i.e., node $2$).
Note that node $1$ broadcasts its updated state information more frequently than node $3$ since the leader node reacts to the command earlier than node $3$.
Consequently, node $2$ can use this broadcast frequency to identify the leader's direction, increasing the edge weight in that direction more than other edges (i.e., selecting larger $\gamma_{ij}(t)$).
Hence, node $2$ quickly aligns with the leader, while the influence of node $3$ remains limited due to a reduced edge weight with node $3$ (i.e., selecting smaller $\gamma_{ij}(t)$).
This framework extends to the entire network.

We note that if (\ref{inner_triggering}) is not violated for node $i$, all edge weights of node $i$ are reduced to $a_\text{min}$ without any prioritization.
In the absence of an active task (i.e., idle phase), it is desired to minimize the influence of nodes between each other, which also decreases the number of events.

Now, we are ready to present the dynamics of $\gamma_{ij}(t)$.
We first define $\phi_{ij}(t)$, which represents the number of state updates received by node $i$ from its neighboring node $j$. 
Note that a node counts the number of updates received from its neighboring nodes if (\ref{data_recording}) is violated.
Otherwise, it is set to zero ($\phi_{ij}(t)=0)$.
Next, $\gamma_{ij}(t)$ can be defined as
\vspace{0cm}
\begin{eqnarray}
\dot{\gamma}_{ij}(t) =
\begin{cases}
 -\psi(\gamma_{ij}(t)-\underline{{\gamma}_{ij}}), \quad \text{if} \  \phi_{ij}(t)<\phi_{ir}(t),
 \\
-\psi(\gamma_{ij}(t)-\overline{{\gamma}_{ij}}), \quad \text{otherwise}.
\end{cases}
\label{gamma_update_rule}
\end{eqnarray}
In (\ref{gamma_update_rule}), we consider that $j,r\in \mathcal{N}_i$ such that $j\neq r$.
The parameter $\psi\in \IR_{>0}$ denotes the design parameter.
In addition, $\underline{{\gamma}_{ij}}\in \IR_{>0}$ and $\overline{{\gamma}_{ij}}\in \IR_{>0}$ represent lower and upper bounds, respectively ($\overline{{\gamma}_{ij}}>\underline{{\gamma}_{ij}}$).
As noted in Remark 1, these bounds are used to dynamically prioritize edge weights: if node $j$ sends fewer updates to node $i$ compared to another neighbor $r$, the associated control coefficient ($\gamma_{ij}(t)$) is driven toward the lower bound $\underline{{\gamma}_{ij}}$; otherwise, it is adjusted toward the upper bound $\overline{{\gamma}_{ij}}$.

\textbf{Remark 2.} When (\ref{inner_triggering}) is violated, node $i$ initially compares the frequency of state updates (i.e., $\phi_{ij}(t)$).
If any edge weights between node $i$ and its neighboring nodes have been assigned to a lower bound ($\underline{{\gamma}_{ij}}a_{\text{max}}$) in the active phase, those neighbors are excluded from the frequency comparison in (\ref{gamma_update_rule}) until the condition (\ref{outer_event_rule}) is satisfied.\footnote{This exclusion is necessary for the following reason.
Consider that the leader node executes the command and subsequently it reduces the frequency of its state broadcasts.
Over time, it may be possible for some nodes to surpass the leader's broadcast frequency, which leads to information confusion in the network.
Consequently, it increases the number of events and degrades the performance.}

\textbf{Remark 3.} Note that leader node $i$ obtains $a_{ij}(t)$ values from its neighboring nodes instead of using the dynamic edge weight update mechanism.
Specifically, if node $j$ is a neighbor of leader node $i$ and its corresponding edge weight is $a_{ji}(t)$, then the leader node sets $a_{ij}(t)=a_{ji}(t)$.

\textbf{Remark 4.} In addition to leader nodes, some nodes in the network may have only one neighboring node (i.e., node $4$ in Figure \ref{graph0}).
In that case, node $i$ cannot compute its $a_{ij}(t)$ using the dynamic edge weight update mechanism since it requires the comparison of the number of state updates of its neighboring nodes.
In this scenario, node $i$ sets its $a_{ij}(t)$ to the maximum value of its neighbor edge weight value.

Before presenting our main results, we introduce an assumption used in the following section.

\textbf{Assumption 1.} Let $c(t)\in \IR$ be a continuously differentiable time-varying function with a bounded and continuously differentiable derivative $\dot{c}(t)$.


\section{Main Results and Stability Analysis}\label{MR}

Consider the dynamics of the multiagent network given in (\ref{node_dynamics}) and (\ref{control_signal}).
One can write (\ref{node_dynamics}) as
\vspace{0cm}
\begin{eqnarray}
\dot{x}_i(t)\hspace{-0.3cm}&=&\hspace{-0.3cm}-\sum_{j\in \mathcal{N}_i} a_{ij}(t)\bigg((x_i(t)-x_j(t)) + k_i (x_i(t)-c(t))\bigg) \nonumber \\ && \hspace{-0.3cm}-\sum_{j\in \mathcal{N}_i} a_{ij}(t)\bigg((x_{\rom{s}i}\hspace{-0.01cm}(t) -x_i(t)) - (x_{\rom{s}j}\hspace{-0.01cm}(t)-x_j(t))\bigg) \nonumber \\ && \hspace{-0.3cm}-k_i \sum_{j\in \mathcal{N}_i} a_{ij}(t)\left(x_{\rom{s}i}\hspace{-0.01cm}(t)-x_i(t)\right), \ x_{i}(0)=x_{i0}.
\label{state_dynamics_with_control_signal}   
\end{eqnarray}
Now, we define the error 
\begin{eqnarray}
    e_i(t) \triangleq x_i(t)-c(t), \quad \forall  i\in \left\{1,\ldots,n\right\}.
    \label{error_definition}
\end{eqnarray}
The time derivative of the error can be given by 
\vspace{0cm}
\begin{eqnarray}
\dot{e}_i(t)\hspace{-0.2cm}&=&\hspace{-0.2cm}-\sum_{j\in \mathcal{N}_i} a_{ij}(t)\bigg(e_i(t)-e_j(t) + k_i e_i(t)\bigg) \nonumber \\ && \hspace{-0.2cm}+\sum_{j\in \mathcal{N}_i} a_{ij}(t)\bigg(\tilde{x}_i(t) - \tilde{x}_j(t)+k_i\tilde{x}_i(t)\bigg) \nonumber \\ && \hspace{-0.2cm}-\dot{c}(t), \quad e_i(0)=e_{i0},
\label{node_base_error_dynamics}   
\end{eqnarray}
where $\tilde{x}_i(t)\triangleq (x_i(t)-x_{\rom{s}i}\hspace{-0.01cm}(t))$.
Then, one can compactly write (\ref{node_base_error_dynamics}) as
\begin{eqnarray}
    \dot{e}(t)\hspace{-0.2cm} &=&\hspace{-0.2cm} -F(\mathcal{G}(t))e(t)+F(\mathcal{G}(t))\tilde{x}(t)-\textbf{1}_n\dot{c}(t), \nonumber \\  && \hspace{4cm} e(0)=e_0.
    \label{error_dynamics}
\end{eqnarray}
Here, $F(\mathcal{G}(t)) \hspace{-0.06cm} \triangleq \hspace{-0.06cm}  \mathcal{L}(t)+K\mathcal{D}(t)$, where $K\hspace{-0.02cm}\triangleq\hspace{-0.02cm}\text{diag}(k_1\hspace{-0.01cm},\ldots,\hspace{-0.01cm}k_n)$.
Also, $e(t) \hspace{-0.1cm} \triangleq \hspace{-0.1cm} [e_1\hspace{-0.03cm}(t),\ldots,e_n\hspace{-0.03cm}(t)]\mT$ and $\tilde{x}(t) \hspace{-0.1cm}\triangleq \hspace{-0.1cm}  [\tilde{x}_1\hspace{-0.03cm}(t),\ldots,\tilde{x}_n\hspace{-0.03cm}(t)]\mT$\hspace{-0.07cm}.

\textbf{Theorem 1.} Consider a multiagent network composed of $n$ nodes over a time-varying, connected, and undirected graph $\mathcal{G}(t)$ under Assumption 1.
Consider also the dynamics of nodes in (\ref{node_dynamics}), the distributed control architecture in (\ref{control_signal}), the state broadcast event-triggering rule in (\ref{outer_event_rule}), the edge weight update event-triggering rule in (\ref{inner_triggering}) along with dyamic edge weight update mechanism in (\ref{aij_dynamics}), (\ref{thetaij_dynamics}), (\ref{edge_weight_dynamics}), and (\ref{gamma_update_rule}).
Then, the closed-loop error dynamics of the overall network given by (\ref{error_dynamics}) is input-to-state stable with $F(\mathcal{G}(t))\tilde{x}(t)-\textbf{1}_n\dot{c}(t)$ being considered as the input. 

\textit{Proof.}
To show the closed-loop error dynamics of the overall network (\ref{error_dynamics}) being input-to-state stable, we first demonstrate that the unforced network is given by 
\begin{eqnarray}
    \dot{e}(t) = -F(\mathcal{G}(t))e(t), \quad e(0)=e_0.
    \label{exponential_stability}
\end{eqnarray}
is globally exponentially stable.
Note that (\ref{exponential_stability}) is globally exponentially stable by [Theorem 2, \citen{gao2015connections}] if the following conditions hold:
\begin{itemize}
    \item[\textit{i)}] There exists a positive constant $L$ such that for all $t$, $||-F(\mathcal{G}(t))||_2\leq L$,
    \vspace{0.3cm}
    \item[\textit{ii)}] There exists a positive constant $\sigma_s$ such that for all $t$, 
    \begin{eqnarray}
        \text{Re}\bigg\{\lambda_i\bigg(-F(\mathcal{G}(t))\bigg)\bigg\}\leq -\sigma_s \quad \forall i=\{1,2,,\ldots,n\}. \nonumber
    \end{eqnarray}
    By the first two conditions, there exist positive constants $m$ and $\lambda_0$ (which depend only on $L$ and $\sigma_s$) such that for all $t$,
    \begin{eqnarray}
        ||\text{exp}[-F(\mathcal{G}(t))s]||_2
\leq m \hspace{-0.07cm}\ \text{exp}(-\lambda_0 s) \ \text{for all} \  s\geq 0,    \nonumber \end{eqnarray}
\item[\textit{iii)}] If $-F(\mathcal{G}(t))$ is differentiable and there exist scalars $\alpha>0$ and $0<\mu<\frac{\beta_1}{2\beta_2^3}$ such that for all $t$, $T\geq 0$,
\vspace{0cm}
\begin{eqnarray}
    \int_{t}^{t+T} ||-\dot{F}(\mathcal{G}(s))||_2 ds \leq \mu T +\alpha, \nonumber
\end{eqnarray}
where $\beta_1=\frac{1}{2L}$ and $\beta_2 = \frac{m^2}{2\lambda_0}$.
\end{itemize}
Consider the definition of $F(\mathcal{G}(t))$ given after (\ref{error_dynamics}).
The Laplacian matrix is given by $\mathcal{L}(t) = \mathcal{D}(t)-\mathcal{A}(t)$, where $\mathcal{A}(t)$ consists of $a_{ij}(t)$ that are non-negative and upper bounded by $\overline{{\gamma}_{ij}}a_{\text{max}}$.
Here, $\overline{{\gamma}_{ij}}$ and $a_{\text{max}}$ are positive constant values.
Hence,  $\mathcal{A}(t)$ is bounded.
By the definition of the degree matrix, $\mathcal{D}(t)$ is bounded, and thus $\mathcal{L}(t)$ is bounded as well.
Moreover, $K$ is a diagonal matrix and its diagonal elements are composed of either $1$ or $0$.
Finally, there exists a positive constant $L$ such that for all $t$ satisfying the condition $i)$.

\textcolor{black}{Next, one can show that $-F(\mathcal{G}(t))$ satisfies the condition $ii)$.
Specifically, note that sum of each row of $F(\mathcal{G}(t))$ for all $t$ is zero except the rows corresponding to the leader nodes, where their diagonal elements are larger than the sum of absolute value of their off-diagonal elements.
We also note that the network $\mathcal{G}(t)$ is undirected and connected.
Let $t=t^\ast$ be a time instant, where $t^\ast \ge 0$.
From [Lemma 3.3, \citen{Lewis_book}], one can conclude that all eigenvalues of $F(\mathcal{G}(t^\ast))$ are located in the open right-half plane.
Since $t^\ast$ can be any time, then $\text{Re}\bigg\{\lambda_i\bigg(-F(\mathcal{G}(t))\bigg)\bigg\}$ has negative eigenvalues for all $t$.
As a consequence, the condition $ii)$ holds.}

Recall now the definition of $F(\mathcal{G}(t))$.
Note that $\mathcal{A}(t)$ and $\mathcal{D}(t)$ are composed of $a_{ij}(t)$.
It is clear that the dynamics of $a_{ij}(t)$ given in (\ref{aij_dynamics}) is differentiable along with (\ref{thetaij_dynamics}).
Hence, one can conclude that $-F(\mathcal{G}(t))$ is differentiable. 
One can observe that $-\dot{F}(\mathcal{G}(t))$ is also composed of $\dot{a}_{ij}(t)$, which are bounded.
Therefore, left hand-side of the equation given in $iii)$ is bounded.
We highlight that $\mu$ and $T$ are positive constants.
Then, there always exists a scalar $\alpha>0$ satisfying the condition $iii)$.
Finally, this implies that (\ref{exponential_stability}) is globally exponentially stable. 

Consider Assumption 1, $F(\mathcal{G}(t))$ being a bounded function of $t$, and $\tilde{x}(t)$ being bounded by $\delta$ from (\ref{outer_event_rule}).
Note that the right hand-side of (\ref{error_dynamics}) is continuously differentiable and globally Lipschitz in state and input, uniformly in $t$.
Hence, one can conclude that the error dynamics (\ref{error_dynamics}) is input-to-state stable with $F(\mathcal{G}(t))\tilde{x}(t)-\textbf{1}_n\dot{c}(t)$ being considered as the input, by [Lemma 4.6, \citen{khalil}]. \hfill $\blacksquare$

We now present Theorem 2, which shows that the closed-loop network does not exhibit a Zeno behavior with the proposed event-triggering framework.

\textbf{Theorem 2.} Given the dynamics of nodes (\ref{node_dynamics}) with control law (\ref{control_signal}) executing the event-triggering rule (\ref{outer_event_rule}) over a time-varying, connected, and undirected graph $\mathcal{G}(t)$ under Assumption 1, the closed-loop network does not exhibit a Zeno behavior.

\textit{Proof.} According to Theorem 1, $e(t)$ is bounded.
Consider (\ref{error_dynamics}) with Assumption 1, it follows that $\dot{e}(t)$ is bounded.  
Then, one can conclude that $\frac{\text{d}}{\text{d}t}x_i(t)$ is bounded over $t\in [0,\infty)$ by using the definition of $e(t)$ given in (\ref{error_definition}). 
Considering the zero-order-hold operator, $x_{\text{s}i}(t)$ satisfies $x_{\text{s}i}(t) = x_i(t_{di})$ over $t\in [t_{di},t_{(d+1)i})$.
Note that $\frac{\text{d}}{\text{d}t}x_{si}(t)=0$ over $t\in [t_{di},t_{(d+1)i})$.
Following that there exists a constant $\xi_i\in \IR_{>0}$ in the sense that
\begin{eqnarray}
    \frac{\text{d}}{\text{d}t}\bigg|x_i(t)-x_{\text{s}i}(t)\bigg| \leq \bigg|\frac{\text{d}}{\text{d}t}x_i(t)-\frac{\text{d}}{\text{d}t}x_{\text{s}i}(t)\bigg|\leq \xi_i,
    \label{zeno_first_bound}
\end{eqnarray}
holds over $t\in [t_{di},t_{(d+1)i})$.
Next, it can be shown that
\begin{eqnarray}
    \lim_{t \to t_{di}^+} |x_i(t)-x_{\text{s}i}(t)| = 0
    \label{limit_zero}
\end{eqnarray}
Integrating both sides of (\ref{zeno_first_bound}) from \(t_{di}\) to \(t\) and using (\ref{limit_zero}), we can obtain
\vspace{0cm}
\begin{eqnarray}
    |x_i(t)-x_{\text{s}i}(t)|\leq \xi_i(t- t_{di}).
    \label{upper_bound_two}
\end{eqnarray}
Note that an event happens once (\ref{outer_event_rule}) is violated.
Then, there must exists a positive constant $\eta_i>0$ such that an event is triggered at $t=t_{(d+1)i}$.
Hence, the following equation holds 
\begin{eqnarray}
    \lim_{t \to t_{(d+1)i}^-} |x_i(t)-x_{\text{s}i}(t)| = \eta_i.
    \label{limit_nonzero}
\end{eqnarray}
Taking the limit of (\ref{upper_bound_two}) yields
\begin{eqnarray}
     \lim_{t \to t_{(d+1)i}^-}|x_i(t)-x_{\text{s}i}(t)|\leq \xi_i \lim_{t \to t_{(d+1)i}^-}(t- t_{di}).
    \label{upper_bound_three}
\end{eqnarray}
From (\ref{limit_nonzero}) and (\ref{upper_bound_three}), one can obtain $\frac{\eta_i}{\xi_i}\leq t_{(d+1)i}-t_{di}$.
Hence, Zeno behavior never occurs. \hfill $\blacksquare$


\section{Illustrative Numerical Examples}\label{numerical_examples}

\begin{figure}[t!]
\hspace{0cm}
\centerline{\includegraphics[width=7 cm]{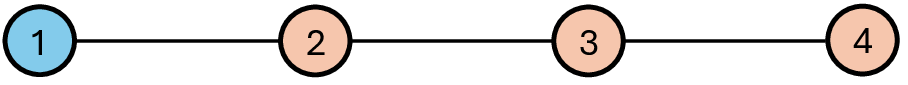}}
\vspace{-0.3cm}
\caption{Multiagent network composed of 4 nodes, where node 1 is the leader and the remaining nodes are followers.}
\vspace{-0.15cm}
\label{graph1}
\end{figure}

\begin{figure}[b!]
\hspace{-0.1cm}
\centerline{\includegraphics[width=8 cm]{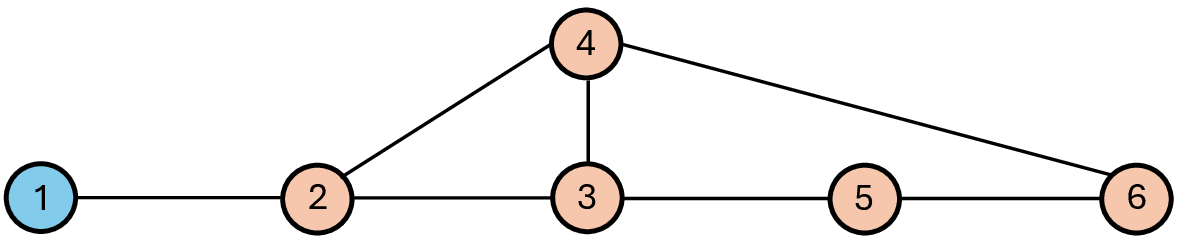}}
\vspace{-0.3cm}
\caption{Multiagent network composed of 6 nodes, where node 1 is the leader and the remaining nodes are followers.}
\vspace{0.0cm}
\label{graph2}
\end{figure}

In this section, we demonstrate the efficacy of the proposed event-triggering framework by comparing it with standard event-triggering method, in which the edge weights $a_{ij}$ between nodes are fixed constant values. 
We consider two multiagent networks as illustrated in Figures \ref{graph1} and \ref{graph2}.
In all examples, a filtered square-wave desired command $c(t)$ is considered.
All initial conditions of nodes (i.e., $x_i(0)$) are selected from the interval $[-1,1]$.
Moreover, all edge weights are initially set to 1 for the proposed method if there exists an edge between nodes $i$ and $j$.
The parameters for dynamics in (\ref{aij_dynamics}) and (\ref{thetaij_dynamics}) are set as $\zeta_1=500$, $\zeta_2=500$, respectively.
In addition, $\psi =500$, $\overline{{\gamma}_{ij}}=1$, and $\underline{{\gamma}_{ij}}=0.3$ are selected\footnote{The parameters can be selected as $\underline{{\gamma}_{ij}}\in [0.2,0.4]$ and $\overline{{\gamma}_{ij}}=1$.} for (\ref{gamma_update_rule}).
Finally, the sampling time is set to $0.001$ sec.

\begin{figure}[b!]
\hspace{2.5cm}
\centerline{\includegraphics[width=13.4 cm]{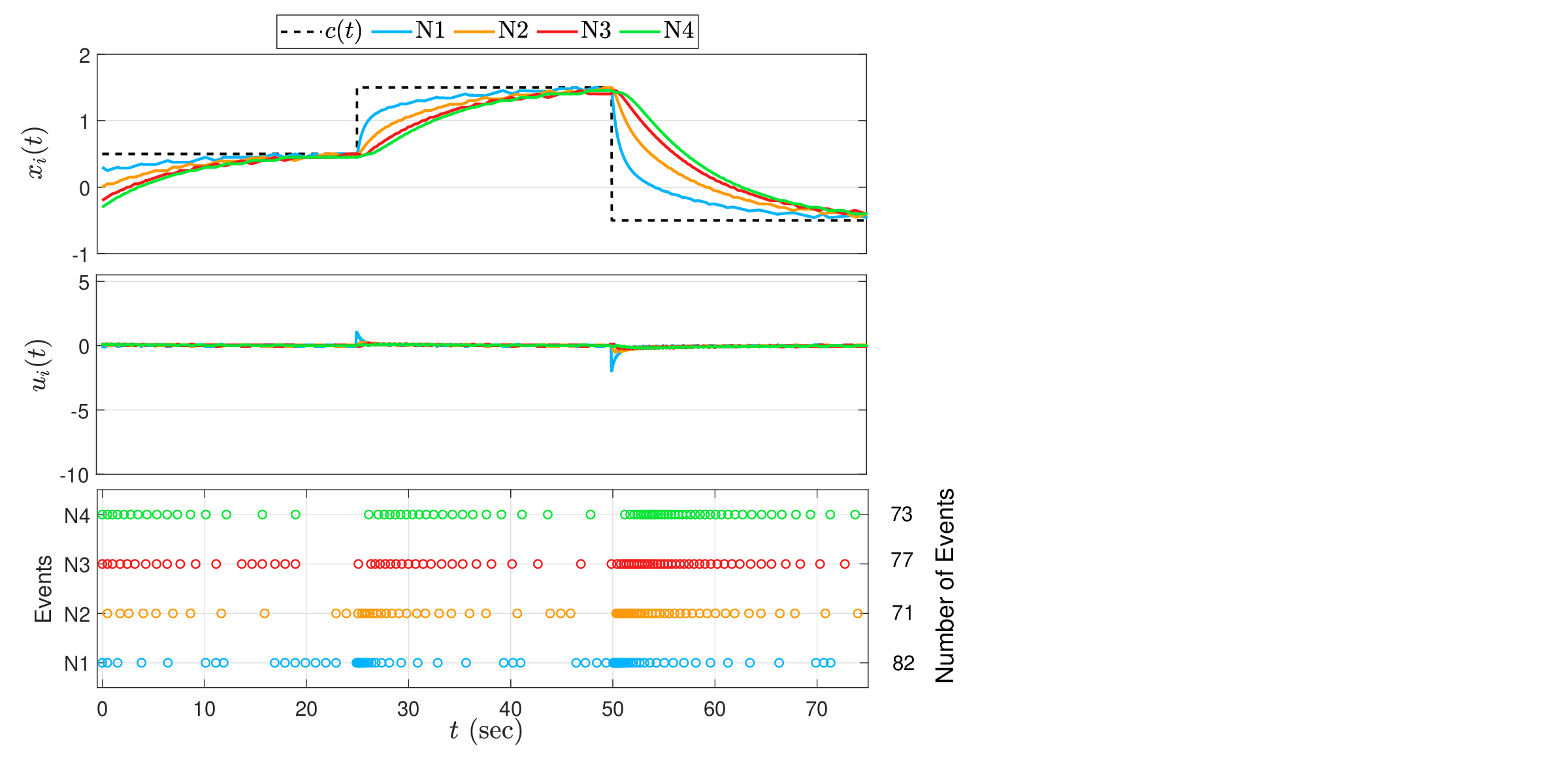}}
\vspace{-0.7cm}
\caption{Illustrative numerical example of the standard event-triggering approach. Edge weights are set to 1. The abbreviation ``N$i$" represents node $i$, $i=\{1,2,3,4\}$.}
\vspace{0.3cm}
\label{example1_standard}
\hspace{2.5cm}
\centerline{\includegraphics[width=13.2 cm]{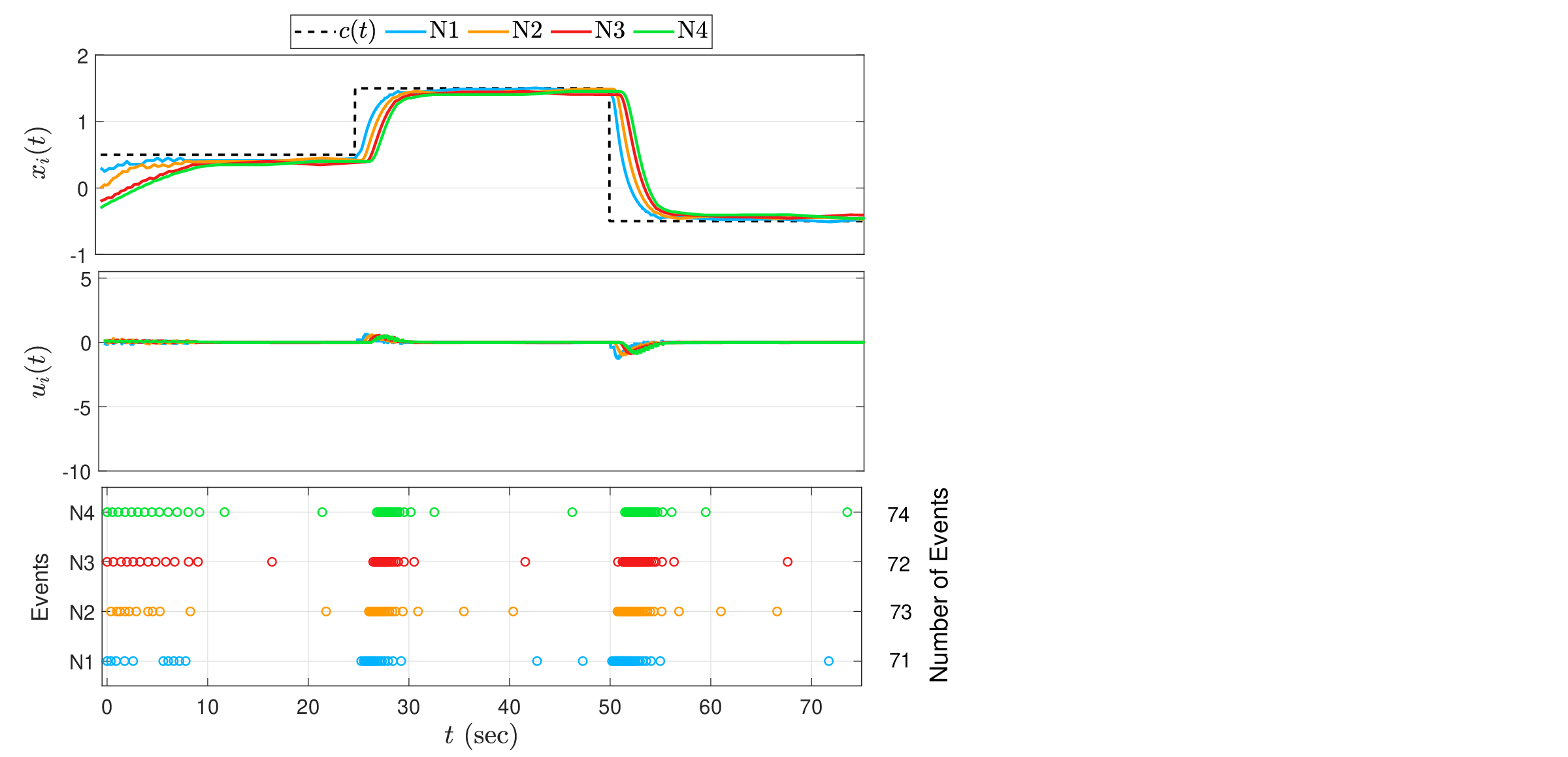}}
\vspace{-0.6cm}
\caption{Illustrative numerical example of the proposed event-triggering framework. Edge weights are time-varying, where $a_{\text{min}}=0.2$ and $a_{\text{max}}=3$. The abbreviation ``N$i$" represents node $i$, $i=\{1,2,3,4\}$.}
\vspace{0.4cm}
\label{example1_proposed}
\hspace{2.75cm}
\centerline{\includegraphics[width=14.4 cm]{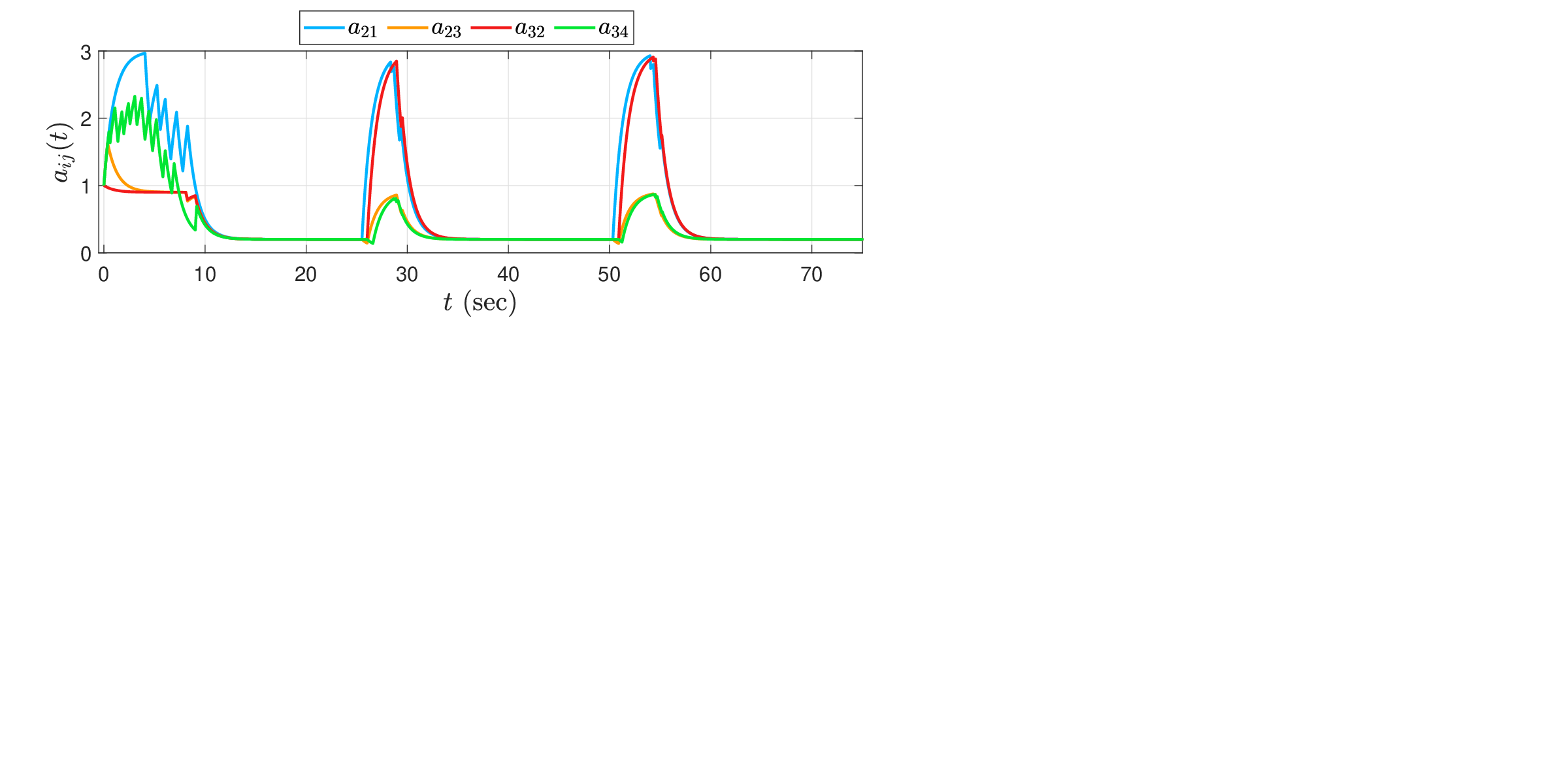}}
\vspace{-4.5cm}
\caption{Changes in edge weights $a_{ij}(t)$ over time for Example 1.}
\label{example1_proposed_edgeweights}
\end{figure}

\textbf{Example 1.} We consider the multiagent network given in Figure \ref{graph1}.
We set $\delta=0.05$ for the event-triggering rule given in (\ref{outer_event_rule}).
Figure \ref{example1_standard} demonstrates the response of the closed-loop multiagent network with standard event-triggering approach, when all edge weights are set to $1$.
Then, we select\footnote{While $a_\text{min}$ should be selected low within $[0.2,1]$ to minimize the interaction between nodes through idle phase, $a_\text{max}$ can be selected within $[2,20]$ to increase the convergence of nodes in the active phase.} $a_{\text{max}}\hspace{-0.1cm}=\hspace{-0.1cm}3$, $a_{\text{min}}\hspace{-0.1cm}=\hspace{-0.1cm}0.2$ for (\ref{edge_weight_dynamics}).
In addition, $\epsilon=0.08$ is picked\footnote{We observe that selecting $\epsilon$ within  $[1.5\delta,2\delta]$ gives better results.} for the edge weight update triggering rule given in (\ref{inner_triggering}).
The response of the closed-loop multiagent network with the proposed event-triggering framework is given in Figure \ref{example1_proposed}.
In addition, Figure \ref{example1_proposed_edgeweights} shows how edge weights change over time with the proposed method.
From Figures \ref{example1_standard} and \ref{example1_proposed}, it is noticeable that the proposed method reduces the convergence (i.e., approach) time of nodes to the command $c(t)$ and provide better performance with fewer number of events comparing to its counterpart.
\begin{figure}[b!]
\vspace{0cm}
\hspace{2.5cm}
\centerline{\includegraphics[width=13.4 cm]{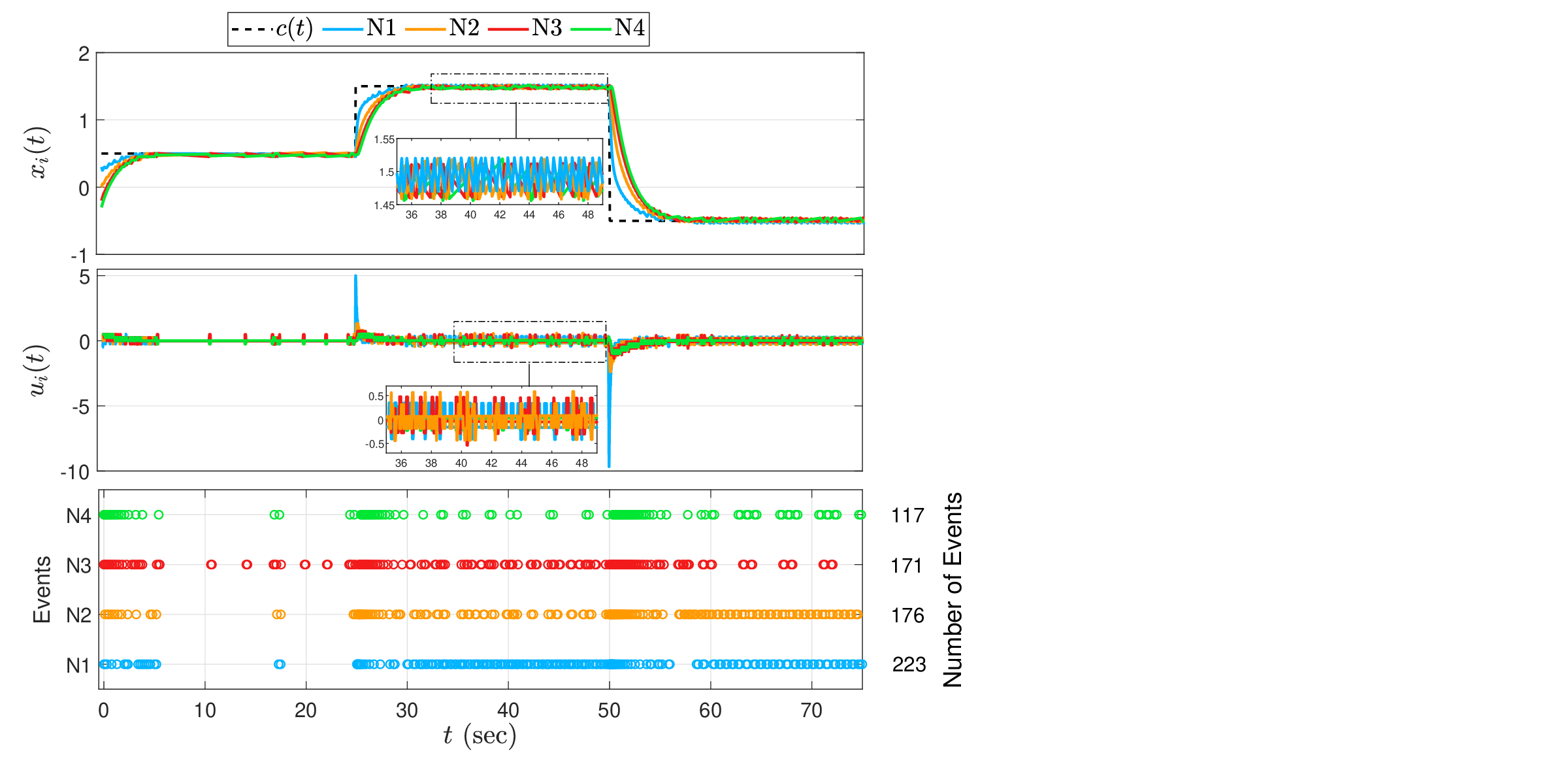}}
\vspace{-0.7cm}
\caption{Illustrative numerical example of the standard event-triggering approach. Edge weights are set to 5. The abbreviation ``N$i$" represents node $i$, $i=\{1,2,3,4\}$.}
\vspace{0.3cm}
\label{example1_standard_highgain}
\hspace{2.5cm}
\centerline{\includegraphics[width=13.2 cm]{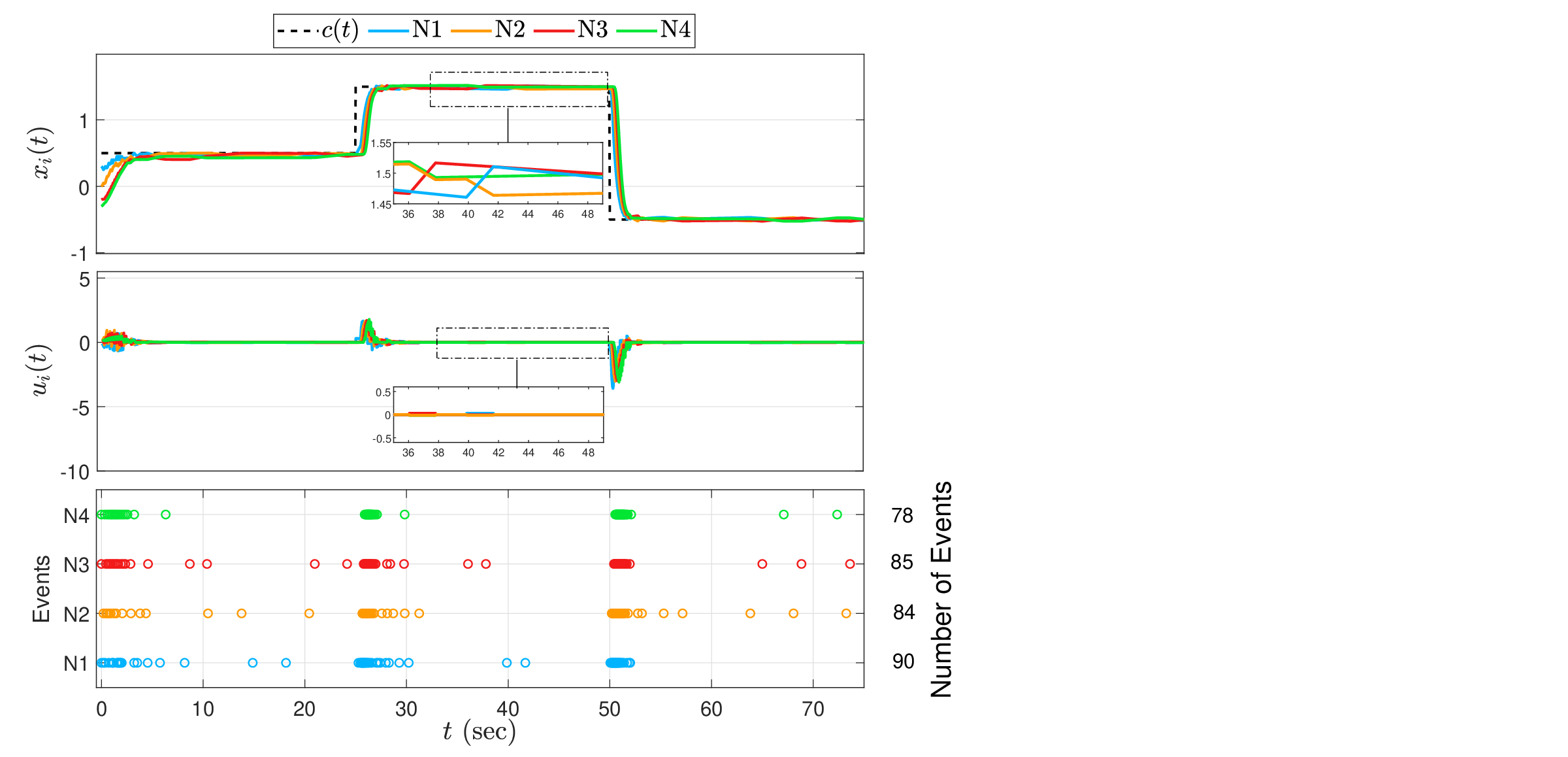}}
\vspace{-0.6cm}
\caption{Illustrative numerical example of the proposed event-triggering framework. Edge weights are time-varying, where $a_{\text{min}}=0.3$ and $a_{\text{max}}=18$. The abbreviation ``N$i$" represents node $i$, $i=\{1,2,3,4\}$.}
\label{example1_proposed_highgain}
\vspace{0.4cm}
\hspace{2.75cm}
\centerline{\includegraphics[width=14.4 cm]{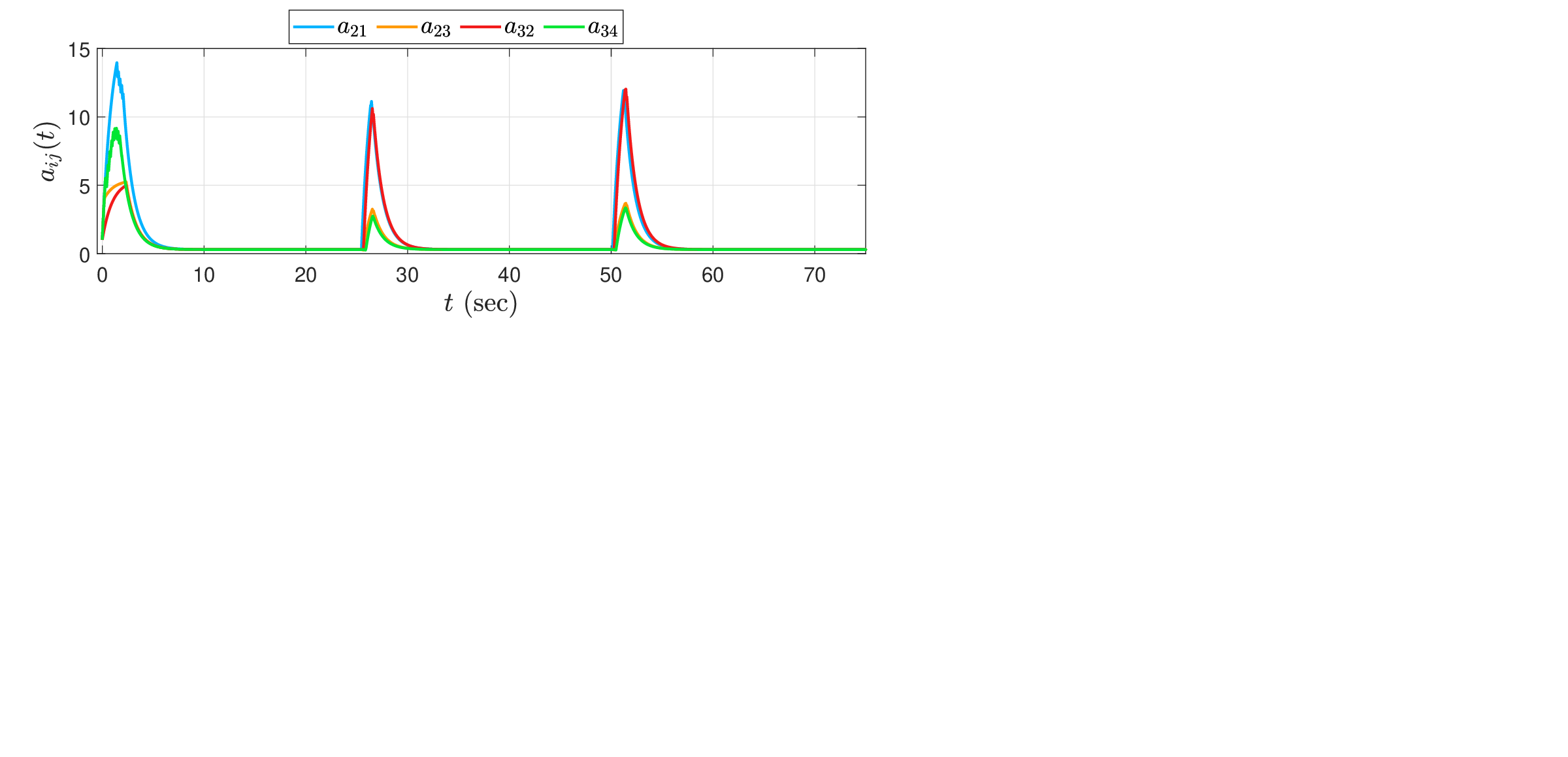}}
\vspace{-4.5cm}
\caption{Changes in edge weights $a_{ij}(t)$ over time for Example 1.}
\vspace{-0cm}
\label{example1_proposed_highgain_edgeweights}
\end{figure}
To better compare the performance of both methods, we show the root-mean-square-error (RMSE) of nodes calculated using the following formula 
\vspace{0cm}
\begin{eqnarray}
    \text{RMSE}= \frac{1}{n}\sum_{i=1}^{n} \sqrt[2]{\frac{1}{M} \sum_{k=1}^M \bigg(x_{\text{s}i}(k)-c(k)\bigg)^2},
\end{eqnarray}
where $n$ and $M$ are the number of nodes and samples in the simulations, respectively.
In addition, we also calculate control effort of nodes by
\vspace{0cm}
\begin{eqnarray}
    \text{E}= \frac{1}{M}\sum_{k=1}^{M} \sum_{i=1}^n ||u_i(k)||_2^2.
\end{eqnarray}
\begin{table}[t!]
\centering
{\small
\caption{Example 1: Comparison of proposed event-triggering control (PETC) with standard event-triggering control (SETC).}
\begin{tabular}{|c|c|c|c|c|}
\hline
         & SETC       & PETC       & SETC        & PETC  \\
         & \makecell{\small$a_{ij}\hspace{-0.05cm}=\hspace{-0.05cm}1$} &  \hspace{-0.15cm}\makecell{\small$a_{ij}\hspace{-0.05cm}\in\hspace{-0.05cm}[0.2,3]$}\hspace{-0.15cm} & \makecell{\small$a_{ij}\hspace{-0.05cm}=\hspace{-0.05cm}5$}  & \hspace{-0.15cm}\makecell{\small$a_{ij}\hspace{-0.05cm}\in\hspace{-0.05cm}[0.3,18]$}\hspace{-0.15cm}     \\
\hline
RMSE     & 0.4998 & 0.3751     & 0.2265      & 0.2314    \\
\hline
 E& 0.2215 & 0.2064 & 0.5540    & 0.2459    \\
\hline
\hspace{-0.2cm} Total Events \hspace{-0.2cm}  & 303    & 290    & 687       & 337     \\
\hline
\end{tabular}
\label{table1}
}
\end{table}
\hspace{-0.15cm}Consistent with our observation, Table \ref{table1} demonstrates that the proposed method achieves lower RMSE, reduced control effort, and fewer total number of  events comparing to the standard approach.


Now, we demonstrate the response of the multiagent network when we select larger $a_{\text{max}}\hspace{-0.1cm}=\hspace{-0.1cm}18$, $a_{\text{min}}\hspace{-0.1cm}=\hspace{-0.1cm}0.3$ for (\ref{edge_weight_dynamics}).
Figure \ref{example1_proposed_highgain} illustrates the response of the network with the proposed framework and Figure \ref{example1_proposed_highgain_edgeweights} shows the change of edge weights.
By comparing Figures \ref{example1_proposed} and \ref{example1_proposed_highgain} that it is evident that larger $a_{\text{max}}$ and $a_{\text{min}}$ lead to improved performance.
However, this enhancement comes with a slight increase in the number of triggering events.
To evaluate the performance of the proposed framework against the standard event-triggering approach, all edge weights in the standard method are fixed to $5$, which yields similar RMSE values as shown in Table \ref{table1}.
A comparison of Figures \ref{example1_standard_highgain} and \ref{example1_proposed_highgain} along with Table \ref{table1} reveals that the proposed framework offers significant advantages over the standard event-triggering approach. Specifically, 
\begin{itemize}
    \item[\textit{i)}]the number of events is reduced by more than half,
    \item[\textit{ii)}] the peak magnitude of control signal is reduced by approximately a factor of three,
    \item[\textit{iii)}] the control effort is reduced by more than half,
    \item[\textit{iv)}]the control signal exhibits noticeably smoother behavior compared to the standard method.  
\end{itemize}

\begin{figure}[h!]
\hspace{2.5cm}
\centerline{\includegraphics[width=13.6 cm]{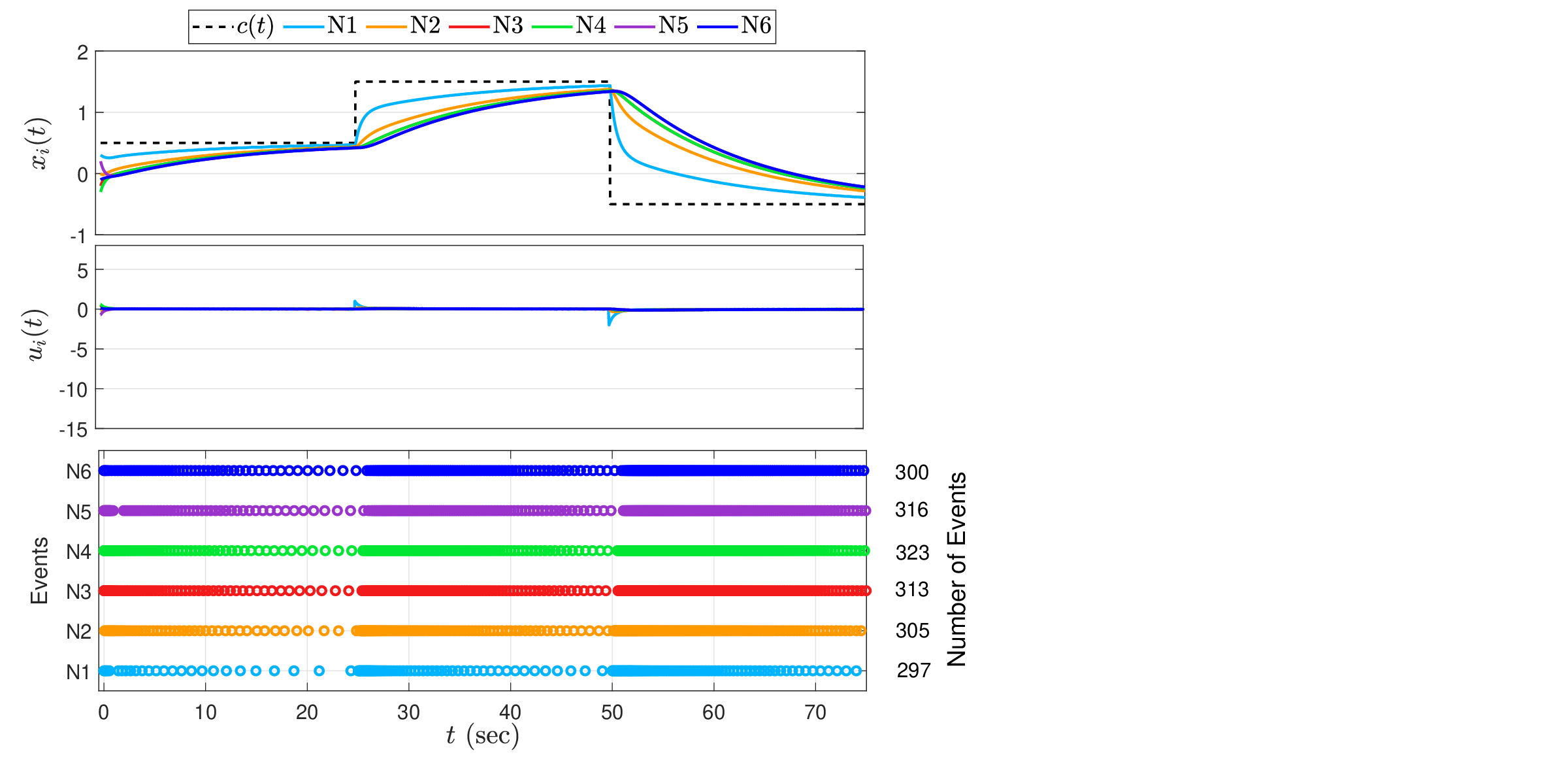}}
\vspace{-0.45cm}
\caption{Illustrative numerical example of the standard event-triggering approach. Edge weights are set to 1. The abbreviation ``N$i$" represents node $i$, $i=\{1,2,3,4,5,6\}$.}
\vspace{0.3cm}
\label{example2_standard}
\hspace{2.5cm}
\centerline{\includegraphics[width=13.25 cm]{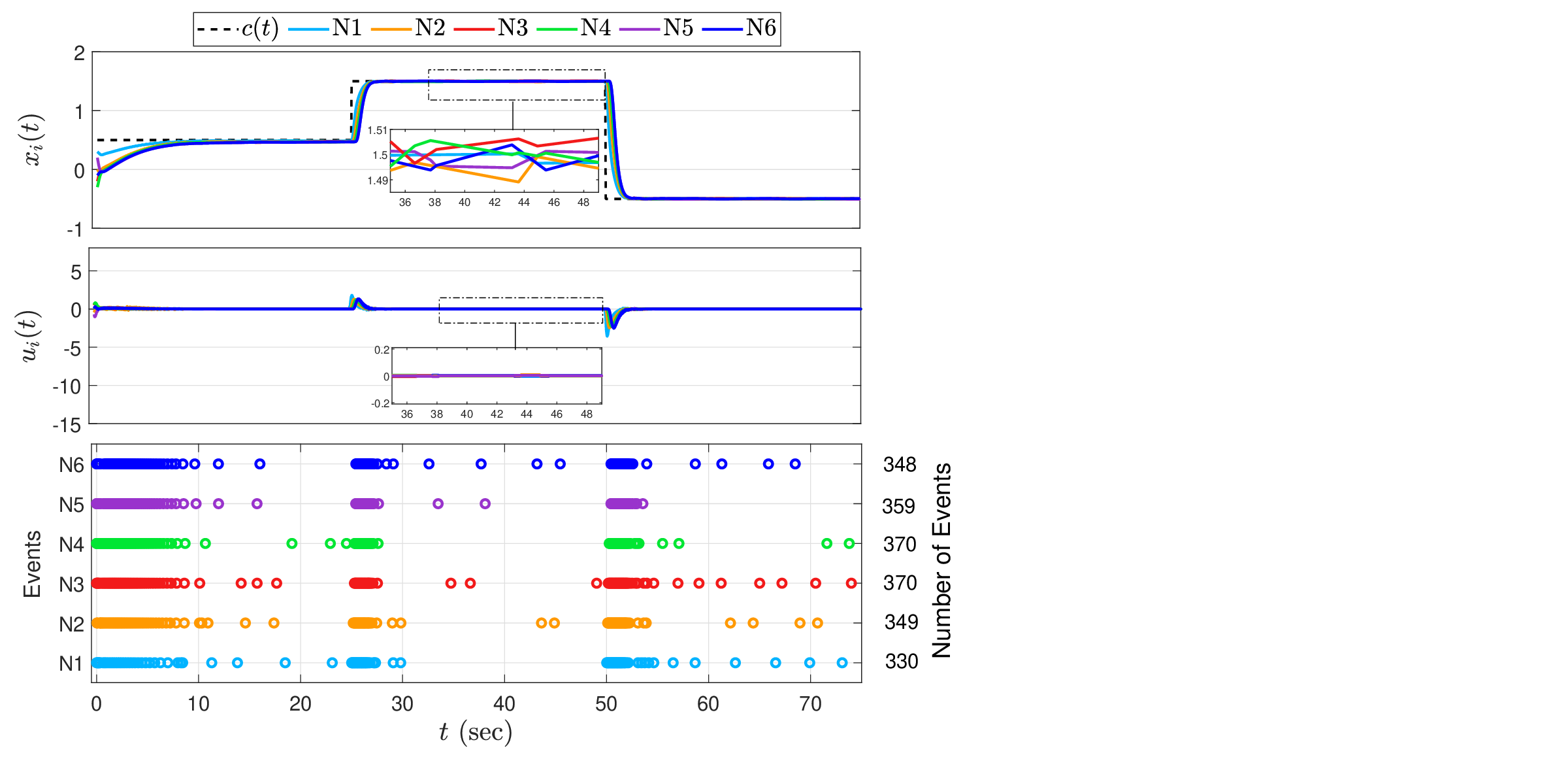}}
\vspace{-0.6cm}
\caption{Illustrative numerical example of the proposed event-triggering framework. Edge weights are time-varying, where $a_{\text{min}}=0.3$ and $a_{\text{max}}=18$. The abbreviation ``N$i$" represents node $i$, $i=\{1,2,3,4,5,6\}$.}
\vspace{0.3cm}
\label{example2_proposed}
\hspace{1.5cm}
\centerline{\includegraphics[width=12 cm]{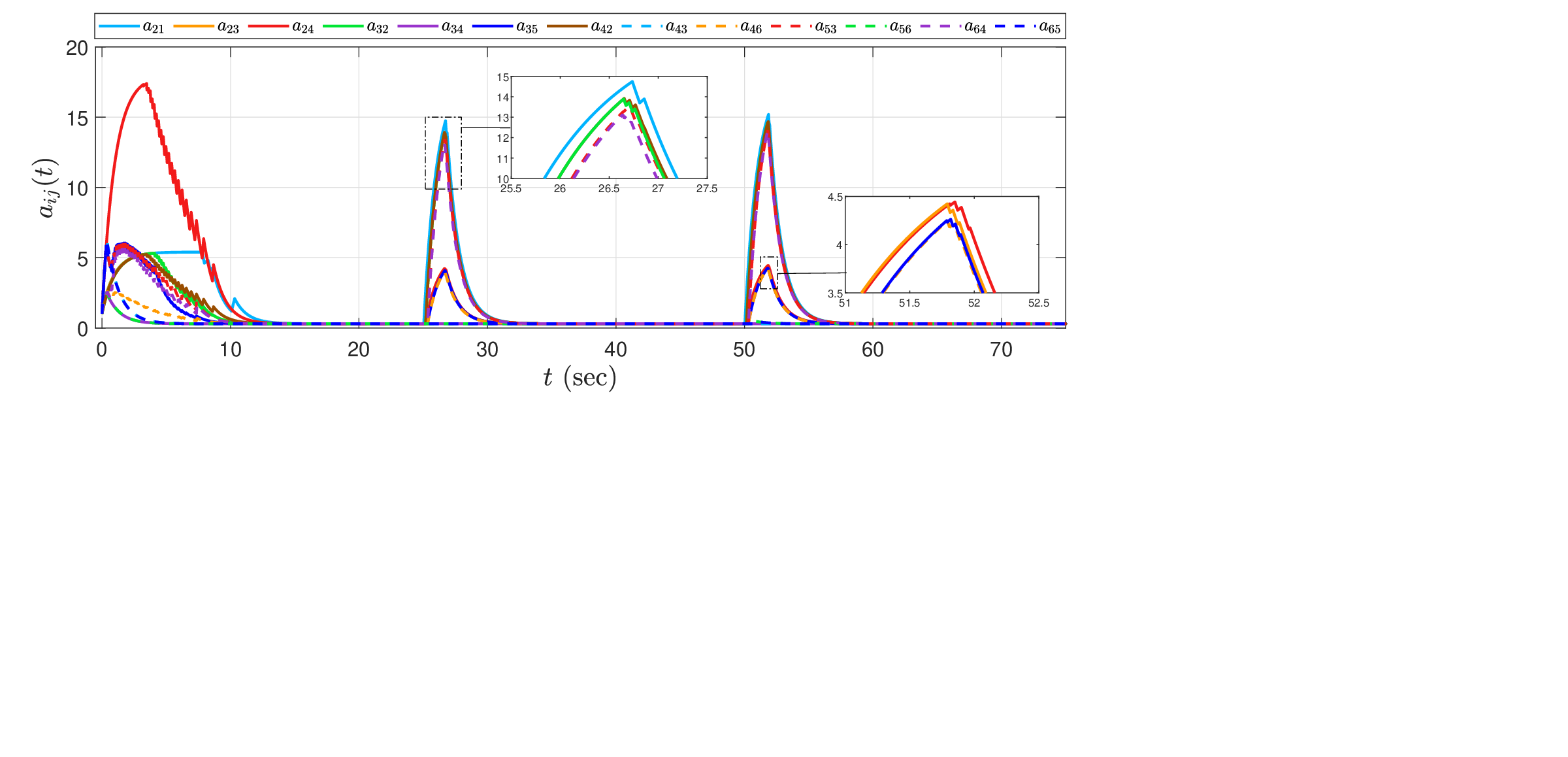}}
\vspace{-3.2cm}
\caption{Changes in edge weights $a_{ij}(t)$ over time for Example 2.}
\label{example2_proposed_edgeweights}
\vspace{-0.5cm}
\end{figure}

\begin{figure}[h]
\hspace{2.5cm}
\centerline{\includegraphics[width=13.5 cm]{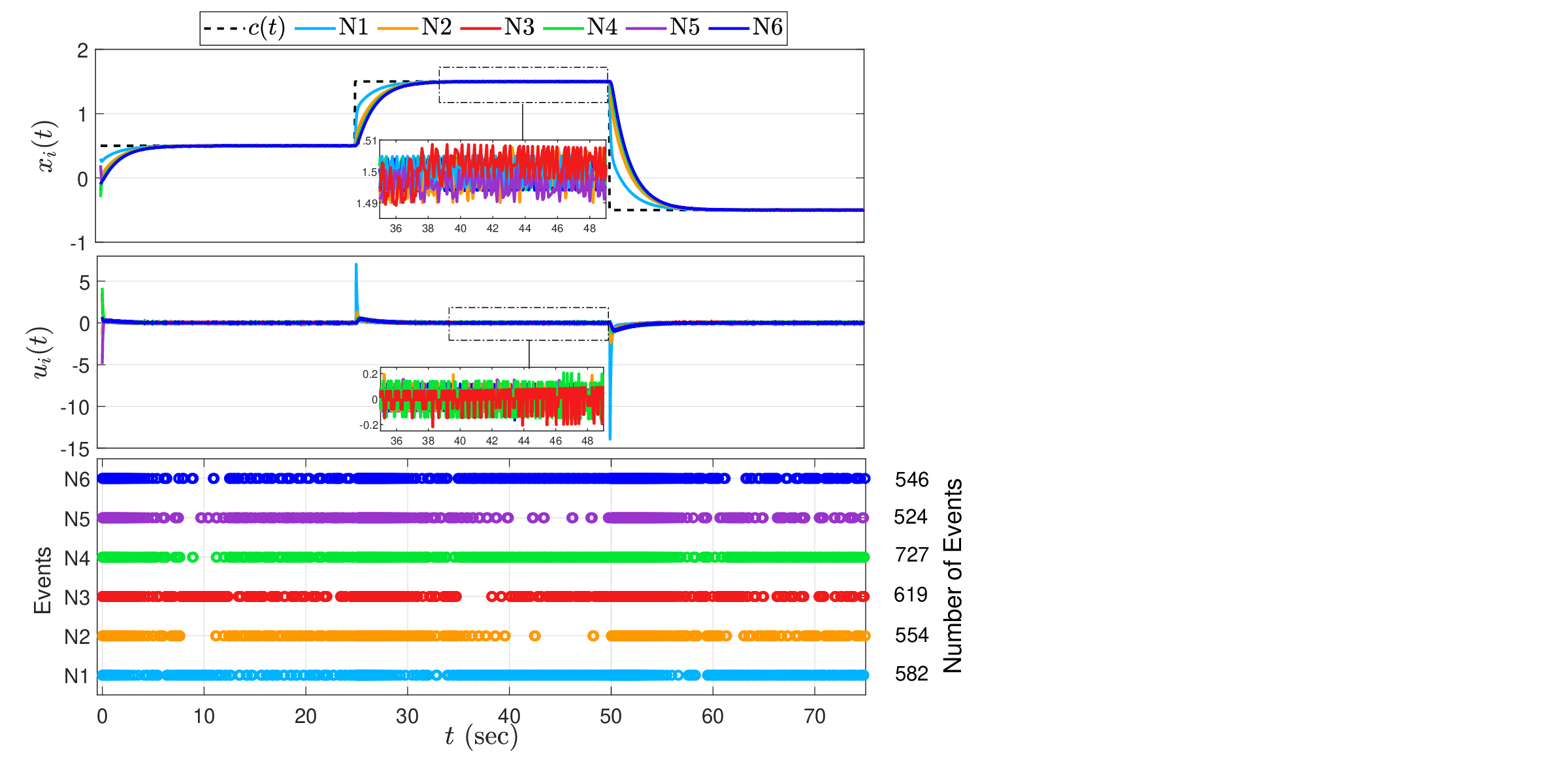}}
\vspace{-0.4cm}
\caption{Illustrative numerical example of the standard event-triggering approach. Edge weights are set to 7. The abbreviation ``N$i$" represents node $i$, $i=\{1,2,3,4\}$.}
\vspace{-0.15cm}
\label{example2_standard_highgain}
\end{figure}

\textbf{Example 2.}  We consider the multiagent network given in Figure \ref{graph2}.
Here, we select $\delta=0.01$ for the event-triggering rule given in (\ref{outer_event_rule}).
Figure  \ref{example2_standard} shows the response of nodes with the standard event-triggering approach when edge weights are set to $1$.
Then, we select $a_{\text{max}}\hspace{-0.05cm}=\hspace{-0.05cm}18$, $a_{\text{min}}\hspace{-0.05cm}=\hspace{-0.05cm}0.3$ for (\ref{edge_weight_dynamics}).
In addition, $\epsilon=0.02$ is picked for the edge weight update triggering rule given in (\ref{inner_triggering}).
Figure \ref{example2_proposed} illustrates the response of the closed-loop multiagent network with the proposed event-triggering framework.
In addition, Figure \ref{example2_proposed_edgeweights} shows how edge weights change over time with the proposed method.
Note that the proposed framework performs better than the standard one with slightly higher number of events and control effort.
To match its performance (similar RMSE values), the standard method requires edge weights set to 7, as shown in Figure \ref{example2_standard_highgain}.
From Table \ref{table2}, Figure \ref{example2_proposed}, and Figure \ref{example2_standard_highgain}, one can again deduce that the proposed framework outperforms the standard event-triggering method, achieving a significantly lower number of events, lower control signal peaks, smoother state and control trajectories. 

\begin{table}[t!]
\centering
{\small
\caption{Example 2: Comparison of proposed event-triggering control (PETC) with standard event-triggering control (SETC).}
\begin{tabular}{|c|c|c|c|}
\hline
 & SETC & SETC & PETC \\
 & \makecell{\small$a_{ij}\hspace{-0.05cm}=\hspace{-0.05cm}1$} & \makecell{\small$a_{ij}\hspace{-0.05cm}=\hspace{-0.05cm}7$} & \makecell{\small$a_{ij}\hspace{-0.05cm}\in\hspace{-0.05cm}[0.3,18]$} \\
\hline
RMSE & 0.6037 & 0.2393 & 0.2254 \\
\hline
E & 0.2515 & 0.5649 & 0.3092 \\
\hline
Total Events & 1854 & 3552 & 2126 \\
\hline
\end{tabular}
\vspace{-0cm}
\label{table2}
}
\end{table}


\section{Conclusion}\label{conclusion}

This paper proposed a time-varying edge weight event-triggered control framework to reduce node-to-node information while improving the performance of the network. 
The key feature of the proposed method is that nodes dynamically adjust their edge weights based on whether the multiagent network is in a transient (active) phase or a steady-state (idle) phase.
Next, we demonstrated the stability of the event-triggered closed-loop multiagent network under this
framework and proved that the closed-loop network does not exhibit a Zeno behavior.
The efficacy of the proposed method was illustrated in Examples 1 and 2 with two different multiagent networks in Section \ref{numerical_examples}. 
As future research directions, edge weight selection can be further improved using optimization techniques and machine learning tools.
In addition, alternative mechanisms such as reputation-based strategies can be employed instead of only using the number of updates received to help nodes prioritize certain edge weights more intelligently. 
Not last but least, the proposed framework could also be extended to event-triggered multiagent networks subject to disturbances and/or uncertainties.






\bibliographystyle{IEEEtran} 
\baselineskip 12pt
\bibliography{refs.bib}
\end{document}